\documentclass[aps,letterpaper,twocolumn,preprintnumbers,floatfix,superscriptaddress]{revtex4}

\usepackage{simplewick}

\usepackage{graphicx}
\usepackage{dcolumn}
\usepackage{bm}
\usepackage{epsfig}
\usepackage{gensymb}
\usepackage{mathrsfs}
\usepackage{amsmath}
\usepackage{amssymb}
\usepackage{graphicx,bm}
\usepackage{slashed}
\usepackage{dsfont}
\usepackage[makeroom]{cancel}
\usepackage{youngtab}
\usepackage{multirow}
 \usepackage[titletoc]{appendix}

\def\fun#1#2{\lower3.6pt\vbox{\baselineskip0pt\lineskip.9pt
  \ialign{$\mathsurround=0pt#1\hfil##\hfil$\crcr#2\crcr\sim\crcr}}}

\newcommand{\Tr}{{\rm Tr\hskip 2pt}}
\def\lsim{\mathrel{\rlap{\raise 2.5pt \hbox{$<$}}\lower 2.5pt\hbox{$\sim$}}}
\def\gsim{\mathrel{\rlap{\raise 2.5pt \hbox{$>$}}\lower 2.5pt\hbox{$\sim$}}}

\input epsf

\newcommand{\be}{\begin{equation}}
\newcommand{\ee}{\end{equation}}
\newcommand{\beq}{\begin{eqnarray}}
\newcommand{\eeq}{\end{eqnarray}}
\newcommand{\bea}{\begin{eqnarray}}
\newcommand{\eea}{\end{eqnarray}}
\newcommand{\ben}{\begin{align}}
\newcommand{\een}{\end{align}}

\usepackage{color}

\newcommand{\comment}[1]{}

\begin{document}

\title{UV Completed Composite Higgs model with heavy composite partners}
\author{Zi-Yu Dong}
\affiliation{CAS Key Laboratory of Theoretical Physics, Institute of Theoretical Physics,
Chinese Academy of Sciences, Beijing 100190, China.}
\affiliation{School of Physical Sciences, University of Chinese Academy of Sciences, Beijing 100190, P. R. China.}
\author{Cong-Sen Guan}
\affiliation{PLA Naval Submarine Academy, Qingdao 266199, China.}
\author{Teng Ma}
\affiliation{Physics Department, Technion -- Israel Institute of Technology, Haifa 3200003, Israel}
\author{Jing Shu}
 \affiliation{CAS Key Laboratory of Theoretical Physics, Institute of Theoretical Physics,
Chinese Academy of Sciences, Beijing 100190, China.}
\affiliation{School of Physical Sciences, University of Chinese Academy of Sciences, Beijing 100190, P. R. China.}
    \affiliation{CAS Center for Excellence in Particle Physics, Beijing 100049, China}
    \affiliation{School of Fundamental Physics and Mathematical Sciences, Hangzhou Institute for Advanced Study, University of Chinese Academy of Sciences, Hangzhou 310024, China}
    \affiliation{Center for High Energy Physics, Peking University, Beijing 100871, China}
    \affiliation{International Center for Theoretical Physics Asia-Pacific, Beijing/Hanzhou, China}
\author{Xiao Xue}
\affiliation{CAS Key Laboratory of Theoretical Physics, Institute of Theoretical Physics,
Chinese Academy of Sciences, Beijing 100190, China.}
\affiliation{School of Physical Sciences, University of Chinese Academy of Sciences, Beijing 100190, P. R. China.}

\begin{abstract}
We study electroweak symmetry breaking in minimal composite Higgs models $SU(4)/Sp(4)$ with purely fermionic UV completions based on a confining Hypercolor gauge group and find that the extra Higgs potential from the underlying preon mass can destruct the correlation between the mass of Higgs and composite partners. Thus the composite  partners can be very heavy for successful electroweak symmetry breaking without enhancing the separation between the new physical scale and Higgs VEV. So this kind of model can be easily realized by ordinary strong dynamics theories without artificial assumptions and more likely consistent with lattice simulations. The UV completion of partial compositeness predicts a light singlet Goldstone boson which interacts  with QCD and electroweak gauge bosons through Wess-Zumino-Witten terms. It can be produced through gluon fusion at LHC and decay into gauge boson pairs. We briefly discuss its phenomenology and derive its bounds from LHC searches.
\end{abstract}

\maketitle

\section{Introduction}
The naturalness of Higgs potential is one of the most profound problems in particle physics. To solve this problem, new physics should be introduced to stabilize Higgs potential. Among these new physics theories, composite Higgs models (CHMs)~\cite{Kaplan:1984plb,Georgi:1984plb,Dugan:1985npb,ArkaniHamed:2002qy} is currently the most popular one. In this model, the Higgs is a composite pseudo-Nambu-Goldstone boson (pNGB) so it is insensitive to other physical scales, such as Planck scale, and thus big hierarchy between electroweak symmetry breaking (EWSB) and the Planck scale can be
achieved.

In ordinary CHMs, the Higgs potential is assumed to be only from top and gauge loop corrections. To regularise Higgs potential and achieve a light Higgs, some composite partners should be introduced to collectively break Higgs shift symmetry or realize maximal symmetry, such as warped extra dimensions~\cite{Randall:1999ee,Contino:2003ve, Agashe:2004rs}, Little Higgs~\cite{ArkaniHamed:2002qy}, and maximal symmetric CHMs~\cite{Csaki:2017cep,Csaki:2018zzf}, which results in strong correlation between the mass of Higgs and composite partners. So there always exists anomalously light top partners, around pNGB decay constant scale $f$, for light Higgs~\cite{Marzocca:2012zn, Matsedonskyi:2012ym,Redi:2012ha}. This special spectrum pattern of composite resonances, very different from QCD (the only observed strong dynamics in nature), requires some artificial
ultraviolet (UV) completions. Moreover, the existing lattice simulations on some confining theories do not support this spectrum pattern~\cite{Bennett:2019jzz,Bennett:2019cxd}, which makes constructing UV completions of ordinary CHMs very challenging.

There is the kind of CHMs that is supposed to have fermionic UV completions based on a confining Hypercolor gauge group $G_{HC}$~\cite{Ferretti:2013kya,Galloway:2010bp,Ryttov:2008xe,Cacciapaglia:2014uja,Ma:2015gra,Csaki:2017jby,Guan:2019qux}. This UV completions contains two species of underlying Weyl fermions called preons, $Q$ (QCD neutral and electroweak charged) and $\chi$ (QCD colored). The confinement of the gauge group $G_{HC}$ will induce the spontaneously global symmetry breaking in the preon sector, generating pNGBs. The doublet pNGBs composed by $Q$ can be treated as Higgs bosons. The colored fermionic bound states with wavefunction $QQ\chi$ or $Q\chi\chi$ can play the role of the top partners, which serves as UV completion of the partial compositeness~\cite{Kaplan:1991dc}. With this setup, there are three types of CHMs with symmetric coset space in EWSB sector: $SU(N_Q)/SO(N_Q)$, $SU(N_Q)/Sp(N_Q)$ and $SU(N_Q/2)^2/SU(N_Q/2)$ ($N_Q$ is the number of chiral preon $Q$), corresponding to $Q$ in the real, pseudo-real and complex representations of $G_{HC}$.

In this work, we study EWSB in the minimal CHMs with global symmetry breaking pattern $SU(4)/Sp(4)\cong
SO(6)/SO(5)$~\cite{Galloway:2010bp,Cacciapaglia:2014uja,Ryttov:2008xe,Csaki:2017jby,Guan:2019qux} in $Q$ sector. If preons $Q$ are massive, Higgs potential will get extra contributions from $Q$ mass naturally. This extra potential can trigger EWSB in a different way together with Higgs potential from top and gauge sector. Especially, the correlation between the mass of Higgs and composite partners is lost (Higgs mass is only related to the scale difference between the partners of top and electroweak gauge bosons) and thus we can get heavy composite partners (can be as heavy as the confine scale $\sim 4\pi f$ at cost of more fine tuning) and light Higgs without enhancing the separation between Higgs VEV and scale $f$.
So this kind of CHMs with heavy fermionic and vector resonances can be easily realized by ordinary strong dynamic theories, such as $G_{HC} =Sp(2N_{HC})$ with $2N_{HC} \leq 36$~\cite{Ferretti:2013kya}, and consistent with lattice simulations, unlike ordinary CHMs.

Besides the extra single scalar $\eta$ in EWSB sector, which is extensively discussed~\cite{Arbey:2015exa,Ma:2017vzm,BuarqueFranzosi:2020baz}, this model predicts another singlet pNGB $\sigma$ associated with $U(1)_\sigma$ global symmetry~\cite{Belyaev:2016ftv}, which is the subgroup of $U(1)_Q$ and $U(1)_\chi$ (overall phase of preon $Q$ and $\chi$). This $U(1)_\sigma$ is anomaly free under $G_{HC}$ so $\sigma$ can be light and crucial for testifying the partial compositeness. This singlet can interact with SM gauge fields (such as gluons) through Wess-Zumino-Witten (WZW) terms. So this singlet can be produced through gluon fusion at LHC and then decay into gauge boson pairs. We briefly discuss its phenomenology at LHC and derive its bounds for different UV completions.

The paper is organized as follows.
In Sec.~\ref{sec:model} we build the concrete UV completions for CHMs based on a confining hypercolor gauge group $G_{HC}$. In Sec.~ \ref{sec:potential}, we calculate the Higgs potential from preon masses,  top and gauge boson loops in two cases: ordinary and minimal maximal symmetric CHMs. In Sec.~\ref{sec:EWSB1}, we study EWSB in Higgs potential and discuss the fine tuning. We find that heavy top partners can be achieved. In Sec.~ \ref{sec:pheno} we discuss the phenomenology of $\sigma$ at LHC and derive its bounds. We conclude in Sec. \ref{sec:conclusion}. The appendices contain
the detailed expressions of top partner multiplets and the form factors in the effective Lagrangian, the descriptions of the gauge sector, as well as the
details of $\eta$ mass from the hidden sector and $\sigma$ mass from $\chi$ sector.
\section{The model}\label{sec:model}
 The consistent UV completions of CHMs with partial compositeness are limited~\cite{Ferretti:2013kya} if satisfy some consistent conditions, such as asymptotically freedom and free of anomalies. In this work, we study  the CHM with global symmetry breaking pattern $SU(N_Q)/Sp(N_Q)$ in $Q$ sector and $SU(N_\chi)/SO(N_\chi)$ in $\chi$ sector. The global symmetry breaking pattern can thus determine that the hypercolor group in the UV completion can only be $G_{HC}=Sp(2N_{HC})$ with $2N_{HC} \leq 36$ or $G_{HC} =SO(N_{HC})$ with $N_{HC} =11,13$~\cite{Ferretti:2013kya}. For simplicity, we focus on the minimal case where $N_Q =4$ and $N_\chi =6$, and the SM custodial symmetry $SU(2)_L \times SU(2)_R \subset SU(4)$ (Hypercharge is embedded in $SU(2)_R$) and QCD $SU(3)_c \subset SU(6)$ are embedded in the global symmetry as
\beq
 SU(4) \supset SU(2)_L \otimes SU(2)_R : && \bf 4 =(2, 2) \nonumber \\
 SU(6) \supset SU(3)_c:  && \bf 6 = 3 \oplus \bar{3}.
\eeq
The basic set up of our model is summarized in Table~\ref{tab:QN}, where we list the SM quantum numbers of the two species of chiral preons (left-handed Weyl fermion): $Q_{1,...,4},\chi_{1,...,6}$. Under this underlying strong dynamics, the global symmetry actually is $U(1)_\chi \times SU(4) \times U(1)_Q \times SU(6)$, $U(1)_{\chi, Q}$ associated with the universal phase of preons $\chi$ ($Q$), and is broken to $Sp(4) \times SO(6)$. One subgroup of the abelian group $U(1)_{\chi} \times U(1)_Q$ has anomaly with hypercolor symmetry and the corresponding pNGB mass is generally at cut-off scale. While the pNGB associated with the anomaly free subgroup $U(1)_\sigma$ of $U(1)_{\chi} \times U(1)_Q$  can be light, which is defined by the following $U(1)_\sigma$ charge assignment of the preons~\cite{Belyaev:2016ftv},
\beq
q_Q=N_\chi T_\chi,\quad q_\chi=-N_Q T_Q,
\eeq
where $N_{Q,\chi}$ is the number of Weyl fermions $Q/\chi$($N_Q =4$ and $N_\chi =6$ in this model) and $T_{Q,\chi}$ is the Dynkin index of HC gauge group representation of $Q/\chi$. So in this model, the total number of light NGBs at a lower energy scale is
\beq
\bf 26 = 1+5+20,
\eeq
 where $\bf 1$ is from $U(1)_\sigma$ breaking, $\bf 5$ from $SU(4)/Sp(4)$, and $\bf 20$ from $SU(6)/SO(6)$.
\begin{table}[t]
\begin{center}
\begin{tabular}{|c|c|c|c|c|c|}
\hline
 & Sp(2N$_{HC}$)/SO(N$_{HC}$) & SU(3)$_c $  & $SU(2)_L \times SU(2)_R$  & U(1)$_\sigma$ \\
\hline
$Q_{1,...,4}$&F/Spin & 1 & (2,1) $\oplus$(1,2) &$q_Q$ \\
\hline
$\chi_{1,...,6}$ & A/F & 3 $\oplus\;\bar{3}$ &1  & $q_\chi$\\
\hline
\end{tabular}
\caption{Quantum numbers of the Weyl preons under the gauge group $G_{HC}\times SU(3)_c\times SU(2)_L\times U(1)_Y$ and global symmetry $U(1)_\sigma$. The hypercharge is $Y=T_R^3+X$ where $X$ is embedded in the unbroken $SO(6)$ with $X=\text{diag}\{2/3,2/3,2/3,-2/3,-2/3,-2/3\}$. The symbols F, A, Spin means fundamental, 2-index antisymmetric, and spinorial representation of $G_{HC}$ respectively.} \label{tab:QN}
\end{center}
\end{table}
Before identifying the quantum number of these NGBs, we should choose consistent condensations of the underlying preons.
Since the condensations in $Q$ ($\chi$) sector are in the anti-symmetric (symmetric) representation of global symmetry $SU(4)$ ($SU(6)$),
 we can choose the condensation of Q and $\chi$ to be SM gauge invariant and in the form~\cite{Cacciapaglia:2015eqa}:
\beq
\Sigma_{Q0} = \left( \begin{array}{cc}
i\sigma_2 & 0 \\
0 & -i\sigma_2
\end{array}   \right ),  \quad  \Sigma_{\chi 0} =\left( \begin{array}{cc}
0 & 1_{3\times 3} \\
1_{3 \times 3} & 0
\end{array} \right ),
\eeq
which will break the global $SU(4)\times U(1)_\sigma$ to $Sp(4)$ in the electroweak sector and $SU(6)\times U(1)_\sigma$ to $SO(6)$ in the $\chi$ sector. So the quantum number of the NGBs under $SU(3)_c \times SU(2)_L \times SU(2)_R \times U(1)_X$ is
\beq
 SU(4)/Sp(4)&:&  \pi_Q =(1,2, 2)_0+ (1,1,1)_0, \nonumber \\
  SU(6)/SO(6)&:& \pi_\chi = (8,1, 1)_0 + (6,1,1)_{\pm \frac{4}{3}}, \nonumber \\
   U(1)_\sigma &:&  \sigma = (1,1,1)_0,
\eeq
where the subscript represents $U(1)_X$ charge which is the subgroup of $SO(6)$ with embedding $X=\text{diag}\{2/3,2/3,2/3,-2/3,-2/3,-2/3\}$.
Since $SU(4)/Sp(4)$ and $SU(6)/SO(6)$ are symmetric coset space, we can define its automorphism map
\beq
 T \to -V T^T V^T \Rightarrow U \to V U^\ast V^T,
\eeq
where $T$ is the broken generators in $SU(4)/Sp(4)$ or $SU(6)/SO(6)$ coset space and $V$ is the VEV of $SU(4)/Sp(4)$ or $SU(6)/SO(6)$. $U$ is Goldstone matrix fields for $SU(4)/Sp(4)$ or $SU(6)/SO(6)$. So the linearly realized sigma field $\Sigma$ and its transformation under global $SU(N)$ symmetry is
\beq
\Sigma\equiv UVU^T=U^2 V \Rightarrow \Sigma \to g\Sigma g^T, \:\; g \in SU(N).
\eeq
The linearly realized sigma in our model be parameterized as:
\beq
U_{Q,\chi} &=& e^{i\Pi_{Q,\chi}}, \quad
\Sigma_{Q,\chi} = U_{Q,\chi}^2 \Sigma_{Q0,\chi 0} \\
 \quad \Pi _Q& =& \cos \phi  \frac{ \sigma}{2f_Q}  \mathds{1}_4  + \frac{\sqrt{2} \pi^{\hat{a}}_Q}{f}  T^{\hat{a}}
\nonumber \\
\Pi _\chi &=&\sin \phi  \frac{ \sigma}{\sqrt{6}f_\chi} \mathds{1}_6 + \frac{\sqrt{2} \pi^{\hat{A}}_\chi}{f_6}  T^{\hat{A}},
\eeq
where $f_{Q,\chi},f,f_6$ are the decay constants of the Goldstone bosons associated with $U(1)_{Q,\chi}$, $SU(4)/Sp(4)$, $SU(6)/SO(6)$,  $T^{\hat{a},\hat{A}}$ are the $SU(4)/Sp(4)$ ($SU(6)/SO(6)$) broken generators with normalization $\mbox{Tr}[T^a T^b] =\delta^{ab}/2$,  $\phi $ parametrizes the direction of anomaly free $U(1)_\sigma$ subgroup of $U(1)_{Q} \times U(1)_\chi$, with value $\tan \phi \equiv  f_\chi q_\chi /(f_Q q_Q)$.~\footnote{The other orthogonal combination of $U(1)_{Q,\chi}$ correspond to $U(1)_{\sigma^\prime}$ which has anomaly with $Sp(2N_{HC})$, and the associated scalar $\sigma^\prime$ can get heavy enough mass from anomaly and $Sp(2N_{HC})$ instanton effects thus decouple from the theory at lower energy.}
Generally, $U(1)_{Q, \chi}$ broken scale $f_{Q,\chi}$ and  $SU(4)$($SU(6)$) broken scale are determined by the underlying dynamics. In the above goldstone matrix, for simplicity, we have chosen proper $U(1)_{Q,\chi}$ charge of underlying preons to fix $f_{Q,\chi}$ as following,
\beq \label{eq:fQfx}
 f_Q =  f , \quad f_\chi =  f_6.
\eeq
Notice that this choice does not effect  EWSB and its phenomenology.
%
%
Next, we will focus on the electroweak sector. The goldstone bosons associated with $SU(4)/Sp(4)$ can be identified as a Higgs doublet $H$ and an extra single $\eta$,
\bea \label{eq:pNGBs}
   (1,2,2)_0 \oplus (1,1,1)_0=H \oplus \eta.
 \eea
After EWSB, only the physical Higgs $h$ and singlet $\eta$ is remained  so, in the unitary gauge, the explicit form of the Goldstone matrix is (since single $\sigma$ is neutral under SM gauge interactions, we can neglect it for simplicity)
\bea \label{eq:U}
U_Q =\left( \begin{array}{ccc}
 (c^\prime+i \frac{\eta}{\pi_Q } s^\prime )  \mathds{1}_2    & i \sigma^2 \frac{h}{\pi_Q} s^\prime  \\
 i \sigma^2 \frac{h}{\pi_Q} s^\prime   & (c^\prime -i \frac{\eta}{\pi_Q } s^\prime)  \mathds{1}_2
\end{array} \right),
\eea
 where $\pi_Q =\sqrt{h^2 +\eta^2}$,  $c^\prime=\cos \left(  \pi_Q /(2 f)\right) $ and $s^\prime=\sin \left(  \pi_Q /(2 f)\right)$, $\sigma^2$ is the second Pauli matrix.

The covariant kinetic term for these pNGBs is
\beq
\mathcal{L}_g =\frac{ f^2}{8}\text{Tr}[ (D_\mu \Sigma_Q)^\dagger D^\mu \Sigma_Q],
\eeq
from which we can extract the mass of $W$ boson if Higgs acquires a VEV (The gauge interactions preserve $\eta$ shift symmetry so the VEV of singlet $\eta$ does not effect EWSB and in the rest of this paper we always assume its VEV is zero)
\beq
m_W^2 =\frac{1}{4}g^2 f^2 \sin^2 \frac{\langle h \rangle}{f}.
\eeq
So we can find the relation between EWSB scale and global symmetry breaking scale $f$
\beq
\xi \equiv s_h^2  =  \frac{v_{SM}^2}{f^2}, \quad v_{SM} =246   \mbox{GeV},
\eeq
where $s_h \equiv \sin\left( \left<h \right> /f \right)$.

In this work, we only focus on the EWSB in minimal CHM with UV completion and its relevant phenomenology, while the colored pNGBs in $\chi$ sector doest not effect EWSB and they are extensively discussed in~\cite{Cacciapaglia:2015eqa}, so we will not discuss them.

\section{Higgs Potential}\label{sec:potential}
In this section, we analyze the Higgs potential and EWSB based on UV completion. In CHMs the potential of pNGBs is generated by interaction terms that explicitly break global symmetry. In ordinary CHMs, it's common that the SM gauge interaction and top Yukawa couplings are the main sources that contribute to pNGB potential. However, in this model, we will add another important contribution from the preon's mass terms to pNGB potential. Which will bring significantly modifications to pNGB potential.

We want to emphasize again that  since the pNGB $\eta$ and $\sigma$ are electroweak (EW) singlet, their VEV does not effect EWSB. Moreover, we will see that the quadratic terms of their potential can be easily kept positive without fine tuning. So without loss of generality, we always choose $\langle \eta \rangle =0 $ and $\langle \sigma \rangle =0 $ in the following discussions.
\subsection{pNGB potential from preon mass terms}
In this model, the underlying preons can have masses, which will result in the pNGBs potential, like quark masses in QCD. In this subsection, we will discuss contributions of the mass of preon $Q$ to the Higgs potential. The most general gauge invariant mass terms of preon $Q$ that also preserve custodial symmetry are
\beq
\mathcal{L}_{mass}=  Q^{T} _i \Sigma_{m_Q} ^{ij} Q_j + h.c.\;,
\eeq
where $\Sigma_{m_Q}$ is mass matrix,
\beq
\Sigma_{m_Q} =\left( \begin{array}{cc}
i m_{Q_1} \sigma_2 & 0 \\
0 & -i m_{Q_2} \sigma_2
\end{array}   \right ).
\eeq
This mass matrix transforms under global symmetry $SU(4)$ as
\beq
\Sigma_{m_Q} \rightarrow g_Q^* \Sigma_{m_Q} g_Q^\dagger,
\eeq
where $g_Q$ is the $SU(4)$ element.
According to global symmetry of the mass terms, the pNGBs potential generated by preon masses can be given by,
\beq \label{eq:Vm}
V_m &=& - C_Q f^3 \mbox{Tr} [\Sigma_{m_Q}.\Sigma_Q ] +h.c. \nonumber \\
    &=& 8C_Q m_Q f^3    \cos\left( \frac{\sigma \cos \phi }{  f } \right)  \cos \left(\frac{\pi_Q}{f} \right) \nonumber \\
    && -8C_Q \Delta_{m_Q}  f^3  \frac{  \eta }{\pi_Q } \sin \frac{\sigma \cos \phi }{f}  \sin \frac{\pi_Q  }{f} ,
\eeq
where we have defined
\beq
m_Q =\frac{ m_{Q_1} + m_{Q_2}}{2}, \quad \Delta_{m_Q} =\frac{ m_{Q_1} - m_{Q_2}}{2}.
\eeq
Notice that  $C_{Q} \sim \langle Q Q \rangle/(16 \pi^2 f^3)$ is an unknown form factor determined by underlying Hypercolor dynamics~\cite{Galloway:2010bp}, which can be positive or negative.
Generally, besides its potential from $Q$ masses, the potential of singlet $\sigma$ can be also from the mass of $\chi$, more details can be found in App.~\ref{app:sigma_mass}.

\subsection{pNGB potential from  Fermion loops}
As ordinary CHMs, the pNGB Higgs can also get the contributions from the top loop.
In this model, the UV completion constrains the top partners to be in the $\bf 6$ or $ \bf 10$ or $\bf 1$ representation of $SU(4)$ and their wave function, and quantum number under $SU(4)\times SU(6)$ is
\beq \label{eq:top_partner}
\psi_1 &=& \chi Q Q \in  (6,6), \quad \psi_2 =\chi \bar{Q} \bar{Q} \in (\bar{6},6),   \nonumber \\
 \psi_3 &=& Q \bar{\chi} \bar{Q}  \in (1,\bar{6} ), \quad
\psi_4 = Q\bar{\chi} \bar{Q} \in (15,\bar{6}).
 \eeq
Notice that since composite partner $\psi_3$ is a global $SU(4)$ singlet, it can not mix with the top doublet. For the most general case, top fields can mix with these top partners at the same time. But in this work, since we just focus on Higgs potential and EWSB in the CHM with UV completion and this kind of mixings do not change the basic property of Higgs potential as in ordinary CHMs, we just work on the case where top quarks only mix with one multiplet of top partners through some specific dynamics so the shift symmetry of $\sigma$ is always unbroken in case. \footnote{$\psi_1$, $\psi_2$ and $\psi_4$ have different $U(1)_\sigma$ charge so if the top fields couple to two of them simultaneously, the $U(1)_\sigma$ symmetry is broken and the potential of $\sigma$ should be proportional to the product of their mixing couplings. If top quarks only mix with one multiplet of top partners, $U(1)_\sigma$ is always preserved and thus top loops do not contribute to $\sigma$ potential.}  In the following discussions we only focus on the simplest case that  top partners are in $\textbf{6}$ representation of $SU(4)$. Actually, the top quark singlet $t_R$ can mix with the top partners in two ways: one is that $t_R$ is embedded in $\textbf{6}$ representation of global $SU(4)$ to mix with the operator $\psi_1$; the other is that $t_R$ is global $SU(4)$ singlet and directly mix with the $Sp(4)$ singlet component of $\psi_1$. These two cases can result in two different types of Higgs potential if maximal symmetry (MS) exists in the composite sector (corresponding to ordinary MS and minimal MS case respectively).~\footnote{MS is the global symmetry in composite sector while the preon mass terms are in the representation of global symmetry $SU(4)$ and explicitly break it. So the mass terms do not influence maximal symmetry.} In the rest of this subsection, we will discuss these two cases.
\subsubsection{ordinary maximal symmetry}
The left-handed fermionic operators $\psi_1$ can be decomposed under unbroken subgroup $Sp(4)\times SU(3)_c  \times U(1)_X$ as
\begin{align}
(6, 6) &=(5,3,2/3) + (5, \bar{3}, -2/3) + (1, 3, 2/3) +(1, \bar{3}, -2/3)\nonumber\\
& \equiv \Psi_{5L} + \Psi_{5R}^c + \Psi_{1L} + \Psi_{1R}^c,
\end{align}
where the superscript $c$ represents charge conjugation. The contents of these multiplets can be found in Appendix~\ref{app:formula}. In order to mix with these partners, top doublet and singlet, $q_L$ and $t_R$,  should be embedded in $\textbf{6}$ of $SU(4)$ and the embeddings can be chosen as
\begin{align}
\Psi_{q_L} & =  \frac{1}{\sqrt{2}} \left( \begin{array}{cc}
0 & Q_{q_L} \\
-Q_{q_L}^T &0
\end{array} \right),  \quad Q_{q_L} =\left(\begin{array}{cc}
t_L & 0 \\
b_L & 0
\end{array} \right), \nonumber\\
\Psi_{t_R}^c & =\frac{t_R^c}{2} \left( \begin{tabular}{c c}
$-  i\sigma_2$  & 0 \\
0 & $ i\sigma_2$  \\
\end{tabular} \right),
\end{align}
where $t_R^c$ is written in the left-handed form. Notice that the $\eta$ shift symmetry is unbroken for these top embeddings.
According to the transformation properties of the fields, the most general mixing terms between the SM fermions and the top partners invariant under $SU(4)$ global symmetry can be obtained
\bea \label{eq:mix}
\mathcal{L}_{mix} &=& -\lambda_L f \mbox{ Tr}[\Psi_{q_L} U_Q\left(   \Psi_{5R}^c + \epsilon_L \Psi_{1R}^c \right) U_Q ^T  ]  \nonumber \\
& & - \lambda_{R} f \mbox{Tr}[\Psi_{t_R}^c U_Q\left(  \Psi_{5L}+ \epsilon_R \Psi_{1L} \right) U_Q ^T  ]   \nonumber \\
 & &- M_5 \mbox{Tr}[ \Psi_{5L} \Sigma_{Q0} \Psi_{5R}^c \Sigma_{Q0}]  \nonumber \\
&& -M_1 \mbox{Tr}[ \Psi_{1L} \Sigma_{Q0} \Psi_{1R}^c \Sigma_{Q0}] +h.c.\;,
\eea
where $\epsilon_{L,R}$  parameterizes the relative differences between the  mixings of different top multiplets with elementary top fields. To reduce the fine tuning in Higgs potential for successful EWSB, the Higgs potential should be finite. To achieve this, we can assume there is a global symmetry $SU(4)$ which different from Higgs shift symmetry in the composite sector, i.e. maximal symmetry, by following requirements,
\bea \label{eq:condition}
\epsilon_{L,R} =1,\; M \equiv M_1 =M_5.
\eea
Under this condition, we can get the ordinary maximal symmetric model, similar to~\cite{Csaki:2017cep,Csaki:2018zzf}. After integrating out the heavy top partners, we can get the top quark effective Lagrangian with the simplest form,
\bea \label{eq:effective_SU4}
\mathcal{L}_{eff} &=&\Pi_0 ^q(p) \mbox{Tr}[  \bar{\Psi}_{q_L} \slashed p  \Psi_{q_L} ]+  \Pi_0 ^t(p) \mbox{Tr}[  \bar{\Psi}_{t_R}^c \slashed p  \Psi_{t_R}^c ] \nonumber \\
&+&  M_1 ^t(p) \mbox{Tr}[\Psi_{q_L}   \Sigma_Q^*  \Psi_{t_R}^c \Sigma_Q^*]+h.c,
\eea
where $\Pi_{0} ^{q,t}$ and $M_{1}^t$ are form factors and their expressions can be found in App.~\ref{app:formula}.
As discussed in~\cite{Csaki:2017cep,Csaki:2018zzf},  we can find that the maximal symmetry can eliminate the Higgs dependent effective kinetic terms of top quarks in lower energy effective Lagrangian and only effective top Yukawa is dependent on Higgs. Since the effective top Yukawa is collectively generated, $M^t_1 \sim \lambda_L \lambda_R f^2 M$, and leading Higgs potential is proportional to top Yukawa coupling square, Higgs potential must be finite. The top mass is easily obtained,
\beq \label{eq:mtsim}
m_t=\frac{  \lambda_L \lambda_R f^2 M }{ \sqrt{2}M_{T_1}  M_{T_2}  } \sin  \frac{2\langle h \rangle}{f},
\eeq
where $M_{T_1}$ and $M_{T_2}$ are top partners mass and their full expressions are listed in  Appendix~\ref{app:formula}.

Now with this effective Lagrange, we can calculate the Coleman-Weinberg potential of Higgs at the full one-loop level with the form
\begin{equation}\label{eq:HiPo}
 V_t (h) = -2N_c \int \frac{d^4 p_E }{(2\pi)^4 } \mbox{log}\left(1 +\frac{ |M_{1}^t|^2}{2p_E ^2 \Pi_{0}^q \Pi_{0}^t }
  \frac{h^2}{\pi^2_Q}\sin^2\frac{2\pi_Q}{f}\right).
\end{equation}
We can expand $V_t(h)$ in top Yukawa coupling $y_t$ up to $\mathcal{O}(y_t^2)$
\beq \label{eq:Vt2}
V_t(h) \simeq \gamma_f  \left( -\sin^2 \frac{\pi_Q}{f} +\sin^4 \frac{\pi_Q}{f} \right)\frac{h^2}{\pi_Q^2} ,
\eeq
where
\begin{equation}\label{eq:gammaf1}
  \gamma_{f} = 4N_c \int \frac{d^4 p_E }{(2\pi)^4 } \frac{ |M_{1}^t|^2}{p_E ^2 \Pi_{0}^q \Pi_{0}^t }.
\end{equation}
It is easy to find that the Higgs potential in the top sector is equivalent to the Higgs potential in ordinary maximal symmetry and the Higgs VEV naturally lies at $\xi =1/2$.

\subsubsection{minimal maximal symmetry}
In this case, $t_R^c$ is a global $SU(4)$ singlet and thus can only mix with top partner singlet $\Psi_{1}$ directly without dressing nonlinear  pNGB matrix $U_Q$ ($\eta$ shift symmetry is still unbroken).   The interactions between top fields  and top partners can be expressed as
\begin{align} \label{eq:mix2}
\mathcal{L}_{mix} =& -\lambda_L f \mbox{ Tr}[\Psi_{q_L} U_Q\left(   \Psi_{5R}^c + \epsilon_L \Psi_{1R}^c \right) U_Q ^T  ]  \nonumber \\
& - \lambda_{R} f t_R^{c } \Tr[ \Psi_{1L} \Sigma_{Q0} ]
 - M_5 \mbox{Tr}[ \Psi_{5L} \Sigma_{Q0} \Psi_{5R}^c \Sigma_{Q0}]  \nonumber \\
 & -M_1 \mbox{Tr}[ \Psi_{1L} \Sigma_{Q0} \Psi_{1R}^c \Sigma_{Q0}] +h.c.\;.
\end{align}
To achieve finite Higgs potential from the above interactions, we impose MS in $\Psi_{5,1}$ sector again by the conditions in Eq.~(\ref{eq:condition}).  After integrate out these heavy partners, the lower energy effective Lagrangian can be obtained,
\bea \label{eq:effective_SU4_1}
\mathcal{L}_{eff} &=&\Pi_0 ^q(p) \mbox{Tr}[  \bar{\Psi}_{q_L} \slashed p  \Psi_{q_L} ]+  \Pi_0 ^t(p) \bar{t}^c_R \slashed p t_R^{c}  \nonumber \\
&&+  M_1 ^t(p) \mbox{Tr}[\Psi_{q_L} \Sigma_Q^* ] t_R^{c }+h.c.
\eea
The top mass can be extracted
\bea\label{eq:mMtopmass}
m_t =\frac{  \sqrt{2}\lambda_L \lambda_R f^2 M }{M_{T_1}  M_{T_2}  } \sin  \frac{\langle h \rangle}{f}.
\eea

The Higgs potential from full one loop is given by
\bea
V_f   = -2N_c \int \frac{d^ 4 p_E }{(2\pi)^4} \mbox{log}\left(1 +   \frac{ 2|M_1^t| ^2}{ p_E ^2 \Pi_0^t  \Pi_0 ^q } \frac{h^2}{\pi^2_Q}\sin ^2 \frac{\pi_Q}{f}  \right).
\eea
In the limit of $\sin(\pi_Q/f)\ll1$, the Higgs potential can be expanded up to quartic order in $\sin(\pi_Q/f)$,
\bea
V_f   \simeq - \gamma_f\frac{h^2}{\pi^2_Q} \sin ^2 \frac{\pi_Q}{f}  +\beta_f\frac{h^4}{\pi^4_Q} \sin ^4 \frac{\pi_Q}{f},
\eea
with
$$\gamma_f = 2N_c \int \frac{d^4 p_E }{(2\pi)^4 } \frac{2 |M_{1}^t|^2}{p_E ^2 \Pi_{0}^q \Pi_{0}^t }, \beta_f =N_c \int \frac{d^4 p_E }{(2\pi)^4 }\left(\frac{2 |M_{1}^t|^2}{p_E ^2 \Pi_{0}^q \Pi_{0}^t } \right)^2.$$

\subsection{Higgs potential in gauge sector}
As for other CHMs, the elementary EW gauge bosons interact with pNGBs through their mixing with composite vector mesons. According to the UV completion, the preons can confine to form vector mesons with wave function and quantum number under $Sp(4)$ as
\bea
 \rho_\mu^a \sim Q^c  T^a \sigma^\mu Q: {\bf 10}  \quad a_\mu^{\hat{a}} \sim Q^c  T^{\hat{a}} \sigma^\mu Q: {\bf 5},
 \eea
where $T^{\hat{a}}$($T^a$)  is (un-)broken generators of $SU(4)$. These mesons' interactions with EW gauge boson can be determined by hidden local symmetry (more details can be seen in App.~\ref{app:gauge}).  The effective Lagrange of EW gauge boson can be obtained by integrating out these vector mesons,
 \begin{align*}
\mathcal{L}^{\text{eff}} &= \frac{P_t^{\mu \nu}}{2}\Bigg( g^2 \Pi_0 ^W   W^a _\mu  W^a _\nu + g^{\prime 2} \Pi_0 ^B B_\mu B_\nu   \\
& + g^2\Pi_1 \frac{h^2}{\pi^2_Q} \frac{\sin^2 \frac{\pi_Q}{f}}{4} (W_\mu ^1 W_\nu^1 +W_\mu ^2 W_\nu ^2     )   \\
   & + \Pi_1 \frac{h^2}{\pi^2_Q} \frac{\sin^2 \frac{\pi_Q}{f}}{4} (g^\prime B_\mu - g W_\mu ^3 )(g^\prime B_\nu - g W_\nu ^3 )  \Bigg),
\end{align*}
 where $P_t ^{\mu \nu} = g^{\mu \nu} - p^\mu p^\nu/p^2$ is the transverse projector and the explicit expression of form factors, $\Pi_0 ^{W,B}$ and $\Pi_1$, is shown in App.~\ref{app:gauge}. Using the full one-loop Higgs potential in Eq.~(\ref{eq:g_potential}), we can get the leading Higgs potential by expanding it up to $\sin^2 (\pi_Q/f)$ ($\sin^4 (\pi_Q/f)$ and higher power terms are suppressed by gauge coupling so can be neglected, comparing with Higgs potential in the top sector),
\bea
 V_g \simeq \gamma_g \frac{h^2}{\pi^2_Q} \sin^2\frac{\pi_Q}{f},
\eea
with
 $$
\gamma_g=   \frac{3}{8(4\pi)^2 } \int  dp_E ^2 p_E^2\left[ (\frac{3}{\Pi_0 ^W}  +\frac{1}{\Pi_0 ^B} ) \Pi_1\right] .
$$
As QCD, the Higgs potential from gauge loop correct automatically satisfies Weinberg sum rules for CHMs based on fermionic UV completion. So Higgs potential is  finite and the leading order of Higgs potential from electroweak gauge bosons loops after imposing Weinberg sum rules is~\cite{Marzocca:2012zn,Csaki:2017jby}
\bea \label{eq:Vg}
V_g \simeq \frac{3f^2(3g^2+g^{\prime2})m_\rho^2\ln2}{64\pi^2}\frac{h^2}{\pi^2_Q}  \sin ^2\frac{\pi_Q}{f},
\eea
where for simplicity we require the scale $f_\rho$ associated with vector meson mixing with SM gauge boson to be equal to $f$, $f_\rho =f$.
\section{Analysis of the Higgs potential}
\label{sec:EWSB1}
In this section, we will discuss EWSB, the spectrum of new fields, and fine tuning in Higgs potential. We will find that the Higgs potential from preon $Q$ mass can weaken the correlation between Higgs mass and top partner mass. So in the CHMs with massive underlying preons, the composite partners can be as heavy as cut-off scale $\sim 4 \pi f$ for successful EWSB.
\subsection{EWSB in Higgs Potential}
\label{sec:EWSB}
The total Higgs potential that determines the EWSB vacuum can be expressed as
\beq
\label{eq:vtheta}
V(h)=   -\gamma s_h^2 +\beta_f s_h^4+ \gamma_m c_h,
\eeq
where $c_h \equiv \cos( \langle h \rangle /f )$, $s_h \equiv \sin( \langle h \rangle /f )$, $\gamma \equiv \gamma_f -\gamma_g $ and $\gamma_m \equiv 8C_m m_Q f^3$ parametrizes Higgs potential from preon masses. In the ordinary MS case, $\beta_f =\gamma_f$. Notice that we always assume that the VEV of singlet $\sigma$ and $\eta$ is zero for simplicity so the terms in pNGB potential proportional to $\eta$ and $\sigma$ can not effect Higgs VEV and can be eliminated. We will find that this condition can be easily satisfied without fine tuning.  The minimum of the potential that can realize EWSB vacuum is one of the roots of the following equation
\beq \label{eq:vacuum}
\gamma_m+2c_h[\gamma-2\beta_f \xi]=0.
\eeq
If $\beta_f \xi  \ll \gamma$, Higgs vacuum can be estimated as
\bea \label{eq:xi}
c_h \approx -\frac{\gamma_m}{2\gamma}\; \Rightarrow \; \xi \approx  \frac{4\gamma^2 -\gamma^2_m}{4\gamma^2}.
\eea

From this expression, we can find that, different from ordinary CHMs, the EWSB can also be triggered by preon mass contributions.
The mass of Higgs can be extracted from Higgs potential,
\bea
\label{eq:higgsmass}
  m_h^2  =\frac{2\xi[\gamma+2(2-3\xi)\beta_f]}{f^2}.
  \eea
Comparing with ordinary CHMs, the Higgs mass contains extra factor $\gamma$ ($m_h^2 =8\xi \beta_f/f^2 $ in ordinary CHMs). If impose some cancellation between $\gamma$ and $\beta_f$, we can thus easily get the light Higgs and heavy composite partners simultaneously (generally $\gamma$  and $\beta_f$ can have opposite sign and are independent). However, in ordinary CHMs, since Higgs mass is only proportional to $\beta_f$, Higgs mass is strongly correlated with top partner mass and there is no space to tune the parameters to achieve light Higgs and heavy partners simultaneously~\cite{Matsedonskyi:2012ym,Marzocca:2012zn}. So the extra Higgs potential from preon masses can weaken the correlation between Higgs mass and partner mass, which can be explicitly seen in the next subsection.
In our model, the $\eta$ potential is only from the preon mass sector (the gauge and top sector preserve $\eta$ and $\sigma$ shift symmetry), and after EWSB, its mass can be expressed as (for $\xi \ll 1$)
 \bea  \label{eq:eta_mass}
  m_\eta^2  \simeq \frac{f^2m_h^2-8\xi \beta_f}{f^2 \xi}.
\eea
Since $\beta_f$ is positive, to prevent $\eta$ from getting a VEV $\beta_f$ should satisfy the upper limit of $\beta_f < f^2 m_h^2/(8\xi)$, which will impose an upper bound on top partner mass.   However, $\eta$ can obtain extra masses from some terms that only explicitly break $\eta$ shift symmetry so the top partners can be very heavy without violating $\eta$ mass bounds (more details can be seen in App.~\ref{app:eta_mass}).   Notice that for simplicity we assume $\Delta_{m_Q} =0$ such that there is no mixing between $\eta $ and $\sigma$.

 The singlet $\sigma$ both contains the freedoms of underlying preon $Q$ and $\chi$,  its mass can be from both $Q$ and $\chi$ sector~\cite{Belyaev:2016ftv}.
In the $Q$ sector, its mass is only from the mass of preon $Q$ and can be easily extracted in the EWSB phase,
\bea \label{eq:simgamass1}
m_\sigma^{Q} =m_\eta \sqrt{(1-\xi)}\cos \phi.
\eea
Its mass from $\chi$ sector is also generated via the mass of preon $\chi$ (gauge interactions also preserve $\sigma $ shift symmetry in $\chi$ sector), more details can be seen in App.~\ref{app:sigma_mass}. In the rest of this section, we will numerically  calculate the spectrum of the new fields and fine tuning of Higgs potential in two different models.
\subsection{Ordinary Maximal Symmetry}
 In ordinary MS model,  according to the analytical expressions of Higgs potential in Eq.~(\ref{eq:gammaf1}) and Eq.~(\ref{eq:Vg}), the Higgs potential from top and gauge sector are sensitive to the composite partners mass and can be generally parametrized as,
 \bea \label{eq:gamma}
\gamma_f = \beta_f\simeq c_f\frac{N_cy_t^2f^2M_f^2}{8\pi^2}, \quad \gamma_g  \simeq c_g \frac{3g^2 m_\rho^2 f^2}{16 \pi^2},
 \eea
 where $y_t$ is the top Yukawa coupling, $c_{f,g}$ is order one positive parameter, whose analytical expressions can be derived in Eq.~(\ref{eq:gammaf1}) and Eq.~(\ref{eq:Vg}), and $M_f$ is the top partner mass scale. As discussed above, the correlation between Higgs and top partner mass is weakened by the extra Higgs potential from preon mass. Substitute these expressions into Higgs mass in Eq.~(\ref{eq:higgsmass}), we can  see that  Higgs mass is sensitive to the mass scale difference between top and gauge bosons partners,
 \bea
 m_h^2 \sim (5 c_f M_f^2 - c_g m_\rho^2) \xi.
 \eea
 Since $c_{f,g}$ are positive, the light Higgs only indicates that the scale difference between $M_f$ and $m_\rho$ is small while the masses of each composite partner can be very heavy without increasing the scale $f$, just only at the cost of increasing fine tuning. Different from ordinary CHMs, where the Higgs mass is proportional to the mass scale of top partners, $m_h^2 \sim M_f^2\xi$, so the top partners always be light for light Higgs, around $M_f \approx f $, no matter how to tune the parameters if $\xi$ is fixed. For example, in ordinary CHMs based on deconstruction, the maximal value of lightest top partner mass is around $1.5$ TeV for $\xi =0.1$ and $m_h =125$ GeV~\cite{Matsedonskyi:2012ym}. While, in our models, the mass of the lightest top partner can be as heavy as the cut-off ($\sim 4\pi f$) for the same benchmark point if the singlet $\eta$ can acquire extra mass through some hidden interactions that only explicitly break its shift symmetry, as shown in App.~\ref{app:eta_mass}. As discussed in Sec.~\ref{sec:EWSB}, if $\eta$ potential is only from preon mass, $\eta$ mass is correlated with top partners mass, which imposes the upper bound on top partners. For example if $\xi =0.05$ and $m_\eta^2 >0$, using the expressions in Eq.~(\ref{eq:gamma}), top partner mass $M_f$ should satisfy
 \bea \label{eq:Mf_limit}
 M_f \lesssim \frac{\pi m_h}{y_t \sqrt{c_f N_c \xi}  }  \sim 1.6\, \mbox{TeV}.
 \eea

Next, we will discuss the fine tuning in Higgs potential. Following the convention in~\cite{Barbieri:1987fn}, the fine tuning can be quantified as follows
\begin{equation} \label{eq:tuning1}
  \Delta=\mbox{max}\{\Delta_i\},\;\;  \mbox{with}\,\Delta_i=\left|\frac{\partial\ln\xi}{\partial\ln x_i}\right|,
\end{equation}
where $x_i$ is the free parameter of the model. Using the equation of Higgs vacuum in Eq.~(\ref{eq:vacuum}), we can get the analytical expression of $\Delta_i$,
\bea \label{eq:tuning}
 \Delta_i=\frac{2x_i}{m_h^2f^2}\left[\sqrt{1-\xi}\frac{\partial\gamma_m}{\partial x_i}+2(1-\xi)\left(\frac{\partial\gamma}{\partial x_i}-2\xi\frac{\partial \beta_f}{\partial x_i}\right)\right].
\eea
If $\xi \beta_f \ll \gamma$, according to the approximate expression of $\xi$ in Eq.~(\ref{eq:xi}), to get small $\xi$ the main tuning is from the cancellation between $\gamma_m$ and $\gamma$, which can be expressed as
\bea
\Delta_{m} =\frac{2 \gamma_m^2}{4\gamma^2 -\gamma_m^2} = \frac{2}{\xi}(1- \xi).
\eea
 Under this condition, the tuning is always minimal even $\gamma_f \gg \beta_f$ which always results in double tuning $\Delta \approx \gamma_f/(\xi \beta_f)\gg 1/\xi$ in ordinary CHMs. So the preon mass can relax double tuning in ordinary CHMs.
In general situation, the tuning is mainly from the cancellations among preon mass, gauge, and top sector. The tuning from these three sectors has the same behavior and almost is at the same order of magnitude. We can explicitly look at the tuning from $\rho$ meson mass through Eq.~(\ref{eq:tuning}),
\bea \label{eq:rho_tuning}
 \Delta_\rho=\left|\frac{\partial\ln\xi}{\partial\ln m_\rho}\right|=\frac{8(1-\xi)\gamma_g}{m_h^2f^2} \sim \frac{m_\rho^2}{m_h^2}.
\eea
If we choose $m_\rho=3$ TeV and fix $\xi=0.05$, we find $\Delta_\rho\sim20$.
Similar to other CHMs, the tuning increases as the mass of composite partners increases. This is because that the Higgs potential is sensitive to the partner mass scale. To get light Higgs, more precise cancellation among $\gamma_{f,g}$ and $\beta_f$ is needed if their masses increase.

\begin{figure}
  \centering
  \includegraphics[width=0.48\columnwidth]{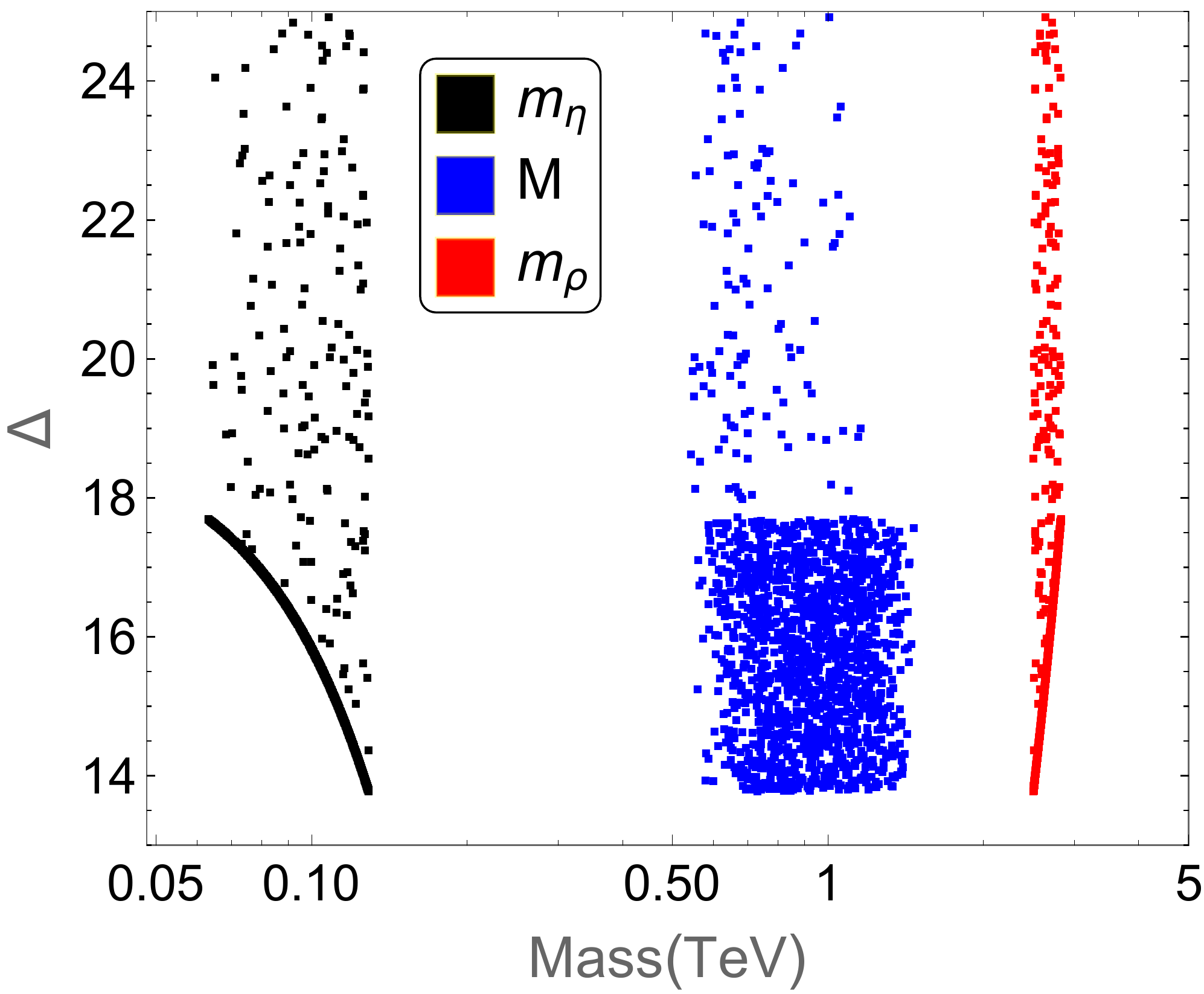}
  \includegraphics[width=0.50\columnwidth]{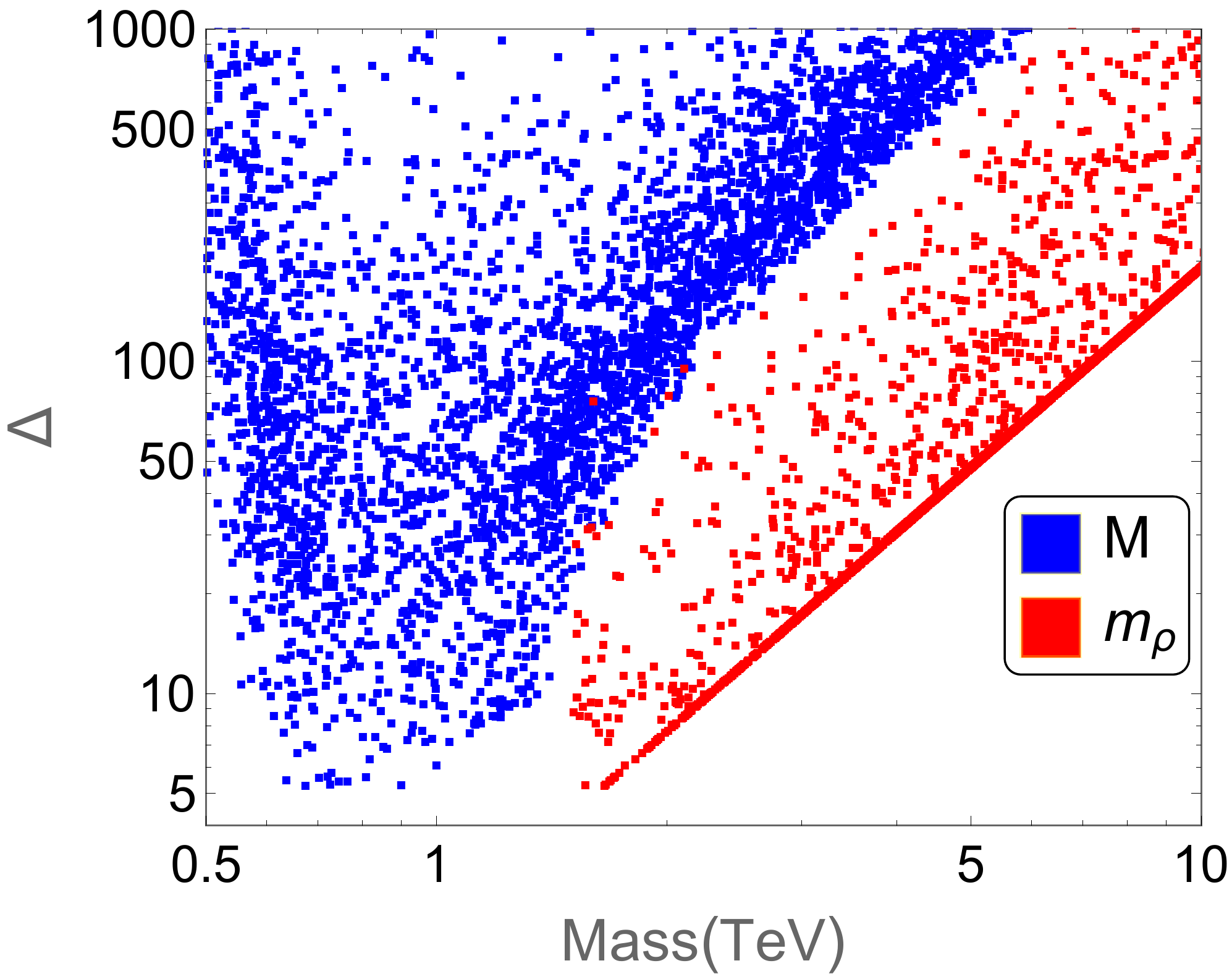}\\
  \caption{Scatter plot of the tuning $\Delta$ in the model with ordinary maximal symmetry as function of mass of $\eta$ (black), lightest top partner (blue) and vector meson $\rho$ (red) for $\xi=0.05$, $m_\eta > m_h/2$, $M >0.5$ TeV and $m_\rho > 2.5$ TeV. The Higgs mass is fixed at 125 GeV and the top mass range is $m_t\in[140,160]$ GeV. Parameter $M$ is the mass of the lightest top partner.}\label{fig:figure1}
\end{figure}

Finally, we use the measurement of fine tuning in Eq.~(\ref{eq:tuning1}) to do the numerical calculations for two cases. One is the minimal case where $\eta$ mass is only from the preon mass sector. In this case, its mass is related to Higgs mass and  top partners mass scale in the EWSB phase.  The other one is that its mass can be also from a hidden sector as shown in App.~{\ref{app:eta_mass}}, so $\eta$ mass can decouple with physics in EWSB. In Fig.~\ref{fig:figure1}, we show the fine tuning as the function of resonance mass for the minimal (left) and non-minimal (right) case with $\xi =0.05$, $m_h =125$ GeV, and $m_t \in [140,160]$ GeV. In the minimal case, since $\eta$ suffers from stringent bounds from Higgs decay~\cite{Khachatryan:2016vau}, we require $m_\eta > m_h/2$ in the numerical scan for consistency. Comparing with the non-minimal case, we can find that, in the minimal case, the bounds of $\eta$ impose an upper limit on lightest top partner mass $M$ (see Eq.~(\ref{eq:Mf_limit})), which also imposes an upper limit on $m_\rho$ thorough Higgs mass. Since $M$ is around scale $f$, the tuning is minimal ($\sim 1/\xi$). In the non-minimal case where $m_\eta$ is not related to $M$, as discussed before, these composite partners can be as heavy as possible for successful EWSB and the tuning increases as these partners become heavy. These numerical results confirm the above analysis.
\subsection{Minimal Maximal Symmetry}
\begin{figure}
  \centering
  \includegraphics[width=0.8\columnwidth]{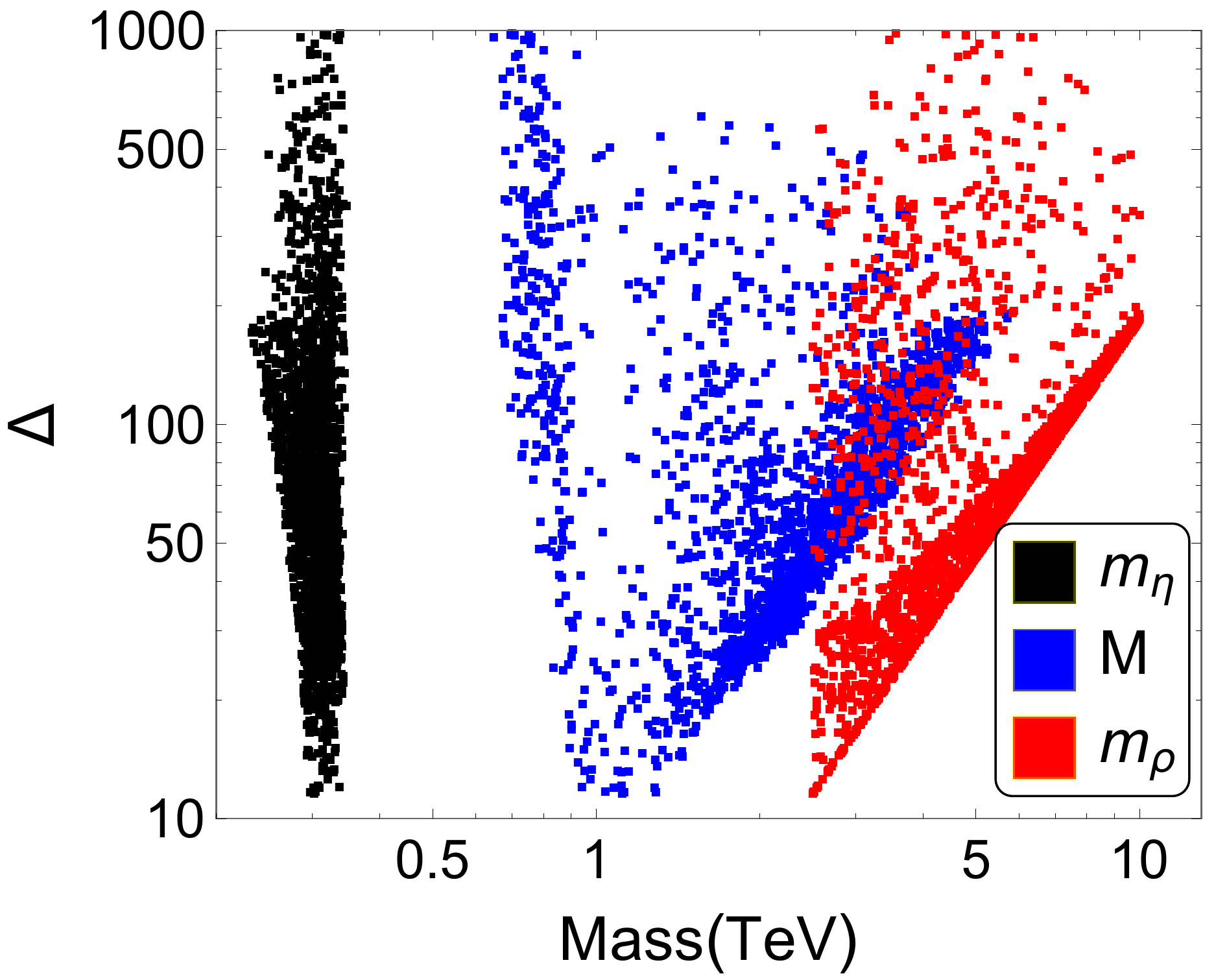}\\
  \caption{Scatter plot of the tuning $\Delta$ in the minimal maximal symmetric model as function of mass of $\eta$ (black), lightest top partner (blue) and vector meson $\rho$ (red) for $\xi=0.1$, $m_\eta > m_h/2$, $M >0.5$ TeV and $m_\rho > 2.5$ TeV. The Higgs mass is fixed at 125 GeV and the top mass range is $m_t\in[140,160]$ GeV.}\label{fig:figure2}
\end{figure}

In the minimal MS CHMs~\cite{Csaki:2018zzf}, the Higgs potential from the gauge sector is the same as ordinary MS CHMs. $\gamma_f$ is also sensitive to the top partner scale, whose parametrization is the same as in Eq.~(\ref{eq:gamma}).  While $\beta_f$ is suppressed at $\mathcal{O}(y_t^4)$ and not sensitive to top partner mass, so Higgs mass is insensitive to $M_f$ in this kind of model. The factor $\beta_f$ can be generally parametrized as
\begin{equation}
 \beta_f\simeq b_f\frac{N_cy_t^4 f^4}{16\pi^2}\ln\frac{M_f^2}{m_t^2},
\end{equation}
where $b_f$ is just order one constant. In this model without preon mass contributions, since $\beta_f$ is suppressed, the Higgs mass is always too light, $m_h \approx 100$ GeV for $M_f \approx 10 f$. Meanwhile, since $\gamma_f$ is much bigger than $\beta_f$, this model suffer from double tuning,  $\Delta \gtrsim 95/\xi$~\cite{Csaki:2018zzf,Csaki:2019coc}. If the preon mass contribution to Higgs potential  is included, the Higgs quartic can be enhanced so Higgs can be heavy enough and the fine tuning can be suppressed ($M_f$ can be reduced). On the other hand, according to the expression of $\eta $ mass in Eq.~(\ref{eq:eta_mass}) in the minimal case, $m_\eta$ is not sensitive to top partner mass and is almost a constant for fixed $\xi$. For example, substituting the expression of $\beta_f$ into Eq.~(\ref{eq:eta_mass}), we can get  $m_\eta \approx 380$ GeV for $\xi =0.1$. Unlike the first model, this model can contain heavy enough $\eta$ to escape the bounds without effecting top partners mass. So the non-minimal case is not necessary.   The behavior of the tuning is the same as the first model. The tuning is minimal for $M_f \sim f$ while it increases as $M_f$ ( $m_\rho$) increases (see Eq.~(\ref{eq:rho_tuning})). In Fig.~\ref{fig:figure2}, we numerically calculate the tuning as the function of resonance masses for $\xi=0.1$ and $m_h =125$ GeV, which confirms the above discussion.

\section{Phenomenology at the LHC}\label{sec:pheno}
If the top partners are very heavy, the first smoking gun of our model may be the presence of  the extra neutral light pNGBs, especially their interactions with SM gauge bosons through WZW terms which can give rise to very typical signatures. We expect that the $\sigma$ field is the first signature of this class of CHMs besides the deviation of the Higgs property, which is the main prediction of partial compositeness, because it has the anomaly interactions with gluon fields, which can result in large production cross-section (the phenomenology of $\eta$ is extensively discussed in~\cite{Arbey:2015exa,Ma:2017vzm,BuarqueFranzosi:2020baz}. Since its production cross-section is very small, its bounds are very weak, $m_\eta > m_h/2$. We will not discuss it in this work).  We will sketch its phenomenology at the LHC in this section  and give some bounds to the parameter space in our model according to the experimental data.

Generally, as discussed before, $\sigma$ can acquire mass from both $Q$ and $\chi$ mass sector,
\bea
m_\sigma =m_\sigma^Q +m_\sigma^\chi,
\eea
where $m_\sigma^{Q,\chi}$ is the mass from $Q$ ($\chi$) sector in Eq.~(\ref{eq:simgamass1}) (Eq.~(\ref{eq:sigmass})).
 If $\sigma$ mass is only from $Q$ sector, it is very light (always light than $m_\eta$ in minimal case) and is excluded by LHC detections. To have heavy $\sigma$, its mass should be dominated by $\chi$ mass contributions in Eq.~(\ref{eq:sigmass}). Since the gauge and top sector preserve its shift symmetry, it does not interact with top or gauge bosons through Yukawa or gauge interactions. However, since $\sigma$ is composed by both $\chi$ and $Q$ freedoms, it mainly interacts with SM gauge fields $A_\mu^i$ through WZW terms, which can be parametrized as follows
\beq\label{eq:WZW}
\mathcal{L}_{WZW}  = \frac{g_i ^2 \kappa_i  }{32\pi ^2 f_\sigma } \sigma \epsilon^{\mu\nu\alpha\beta} A^i _{\mu\nu} A^i _{\alpha\beta },
\eeq
where $f_\sigma \equiv \sqrt{ (q_Q ^2 f_Q ^2 + 3q_{\chi} ^2 f_{\chi} ^2/2 )/(q_{Q}^2  + q_{\chi} ^2 ) } $ is the decay constant associated with $\sigma$, $A_{\mu\nu}^i$ generally denotes the gauge field strength of type $i =W, B,g$ (EW triplet, Hypercharge, gluon). Since $U(1)_\sigma$ is the subgroup of $U(1)_Q$ and $U(1)_\chi$, the coefficients $\kappa_i$ can be obtained from corresponding coefficients of the $U(1)_{Q,\chi}$ Goldstone bosons:
\beq
\kappa_i =\frac{q_Q \kappa_i ^Q +q_\chi \kappa_i ^\chi     }{\sqrt{q_Q ^2  +q_\chi ^2 }},
\eeq
where $\kappa_i^{Q,\chi}$ only depends on the coset space of the $Q$ and $\chi$ condensates, and for our case we have
\bea
\kappa_W ^Q = \kappa_B ^Q = d_Q, \;\;
 \kappa_g ^\chi = 2 d_\chi,  \;\;  \kappa_B ^\chi  = 12 X^2 d_\chi ,
\eea
where $ d_Q /d_\chi $ are the dimension of Hypercolor representation of Q/$\chi$ and $X$ is the $U(1)$ Hypercharge defined in TABLE. \ref{tab:QN}. The main production channel for the $\sigma$ field is through gluon-gluon fusion and the cross-section at proton-proton center-of-mass frame can be parametrized by the partial decay width and the parton luminosities:
\beq
\sigma(pp \to \sigma  ) = \frac{1}{m_\sigma s} C_{gg} \Gamma(\sigma \to gg),
\eeq
where $\Gamma(\sigma \to gg)$ is its decay width to gluon pairs, $s$ is the center-of-mass energy square, the dimensionless partonic integral $C_{gg}$ is
\beq
C_{gg} =\frac{\pi^2}{8} \int^1 _{m_\sigma ^2 /s} \frac{dx}{x} g(x) g(\frac{m_\sigma ^2}{sx} ).
\eeq
The remarkable feature here is that $\sigma$ interactions with gauge fields are completely fixed by the representations of the preons under $G_{HC}$. So $\sigma$ decay widths can reflect the physics of UV completion. The analytical formulae of the partial decay widths to the SM gauge bosons are
\begin{align*}
&\Gamma(\sigma \to gg )  =  \frac{ \alpha_s ^2 \kappa_g^2}{8\pi^3 }  \frac{m_\sigma ^3 }{f_\sigma^2} \\
&\Gamma(\sigma \to \gamma \gamma ) = \frac{\alpha^2 }{64\pi^3 }( \kappa_W + \kappa_B  )^2     \frac{m_\sigma ^3}{f_\sigma^2} \\
& \Gamma(\sigma \to W^+W^- ) = \frac{ \alpha_W ^2 \kappa_W^2}{32\pi^3 } \frac{m_\sigma ^3 }{f_\sigma^2}\left(1-\frac{4m_W ^2}{m_\sigma ^2} \right)^{3/2} \\
&\Gamma(\sigma \to  Z Z ) =
 \frac{\alpha^2 }{64\pi^3 t_W^4}(  \kappa_W +  \kappa_B t_W ^4 )^2    \frac{m_\sigma ^3}{f_\sigma^2}   \left(1-\frac{4m_Z ^2}{m_\sigma ^2} \right)^{\frac{3}{2}} \\
&\Gamma(\sigma \to Z \gamma ) = \frac{\alpha^2 }{32\pi^3 t_W^2 }( \kappa_W   -t_W^2  \kappa_B )^2    \frac{m_\sigma ^3}{f_\sigma^2}  \left (1-\frac{m_Z ^2}{m_\sigma ^2}\right)^{3},
\end{align*}
where $\alpha_W = \alpha/s_W^2$ and $s_W$ ($t_W$) is the sine (tangent) function of Weinberg angle $\theta_W$. Taking the $\sigma \rightarrow \gamma\gamma$ decay channel as the reference channel, we can obtain the ratios of $\sigma$ decay widths for $G_{HC} = SO(11)$, which is only determined by the UV completion and independent on $f_\sigma$:
\begin{equation*}
\frac{\Gamma_{gg}}{\Gamma_{\gamma\gamma}} =870 \;, \frac{\Gamma_{WW}}{\Gamma_{\gamma\gamma}} =4.2 \;, \frac{\Gamma_{ZZ}}{\Gamma_{\gamma\gamma}} =0.416 \;, \frac{\Gamma_{Z\gamma}}{\Gamma_{\gamma\gamma}} =3.1,\;
\end{equation*}
and for  $G_{HC} =SO(13)$:
\begin{equation*}
\frac{\Gamma_{gg}}{\Gamma_{\gamma\gamma}} =711, \; \frac{\Gamma_{WW}}{\Gamma_{\gamma\gamma}} =2.4, \; \frac{\Gamma_{ZZ}}{\Gamma_{\gamma\gamma}} =0.17, \; \frac{\Gamma_{Z\gamma}}{\Gamma_{\gamma\gamma}} =2.3. \;
\end{equation*}
For $G_{HC} =Sp(2N_{HC})$ with $N_{HC} =2$, the ratios are
\begin{equation*}
\frac{\Gamma_{gg}}{\Gamma_{\gamma\gamma}} =68399 \;,   \frac{\Gamma_{WW}}{\Gamma_{\gamma\gamma}} =713.6 \;,\frac{\Gamma_{ZZ}}{\Gamma_{\gamma\gamma}} =144 \;, \frac{\Gamma_{Z\gamma}}{\Gamma_{\gamma\gamma}} =206.2.
\end{equation*}
For the maximum situation, $N_{HC}=18 $, we obtain
\begin{align*}
\frac{\Gamma_{gg}}{\Gamma_{\gamma\gamma}}  &= 76439.5, \;  \frac{\Gamma_{WW}}{\Gamma_{\gamma\gamma}} =1179.7, \\
 \frac{\Gamma_{ZZ}}{\Gamma_{\gamma\gamma}} &= 259.8, \; \frac{\Gamma_{Z\gamma}}{\Gamma_{\gamma\gamma}} =282.4.
\end{align*}
From the above calculations, we can explicitly see that the decay channel  into gluon pairs is  dominant  over the other channels.

\begin{figure}[t]
\begin{center}
\includegraphics[width=0.95\columnwidth]
{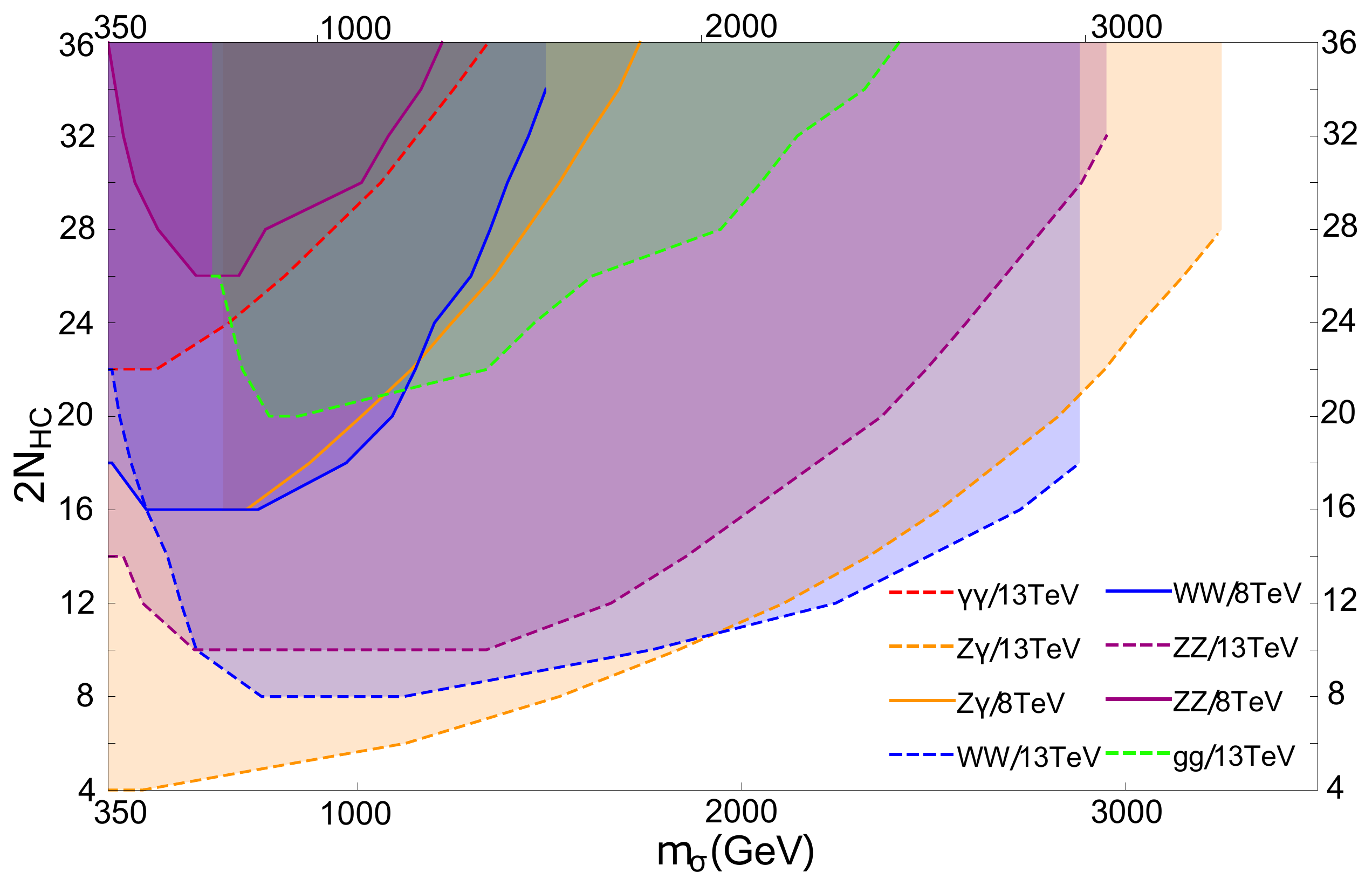}
\end{center}
\caption{Bounds from LHC detections on the mass of $\sigma$ for different hypercolor group $G_{HC}=Sp(2N_{HC})$ with $2N_{HC}\leq36$ and decay constant fixed at $f_\sigma = 800$ GeV. The different color region represents the excluded parameter space via $\sigma$ different decay channels. The cross-sections of the channels with $\gamma\gamma$ and $gg$ final states at $8$ TeV LHC are always smaller than the experimental value and  thus there are no constraints on $m_\sigma$.} \label{fig:SigmaExclusion}
\end{figure}

Using WZW interaction in Eq.~(\ref{eq:WZW}), we simulate different signatures of $\sigma$ in LHC from the following channels,
\bea
g g \to \sigma \to A^i A^j,
\eea
where $A=\{W^\pm, Z, \gamma, g\}$.
By comparing with the experimental data from 8 TeV \cite{aa8,za8,zz8,ww8,gg8} and 13 TeV LHC \cite{ATLAS:WW_13TeV,ATLAS:ZZ_13TeV,ATLAS:gaga_13TeV,CMS:Zga_13TeV,CMS:gg_13TeV}, we derive the bounds of  $m_\sigma$ for different Hypercolor group with $f_\sigma =800$ GeV held fixed as shown in Fig.~\ref{fig:SigmaExclusion} (the color regions are excluded parameter spaces).
For $G_{HC }=Sp(2N_{HC})$ model, the bounds of $\sigma$ increase as $N_{HC} $ increases. For $G_{HC }= SO(11/13)$ hypercolor model, the strongest constraints  come from $Z\gamma$ decay channel and we find $m_\sigma<2.6$ TeV is excluded for $SO(11)$ hypercolor group and $m_\sigma<2.8$ TeV is excluded for $SO(13)$.

\section{Conclusions}\label{sec:conclusion}
We studied the minimal composite Higgs model $SU(4)/Sp(4)$ with purely fermionic UV completions based on a confining Hypercolor gauge group $G_{HC}$. Under this strong dynamics, two species of underlying Weyl fermions in different representations of $G_{HC}$, $Q_{1,..,4}$ (QCD colorless) and $\chi_{1,...,6}$ (QCD colored), should be introduced to generate the composite Higgs doublet, composed by $Q$ alone, as well as top partners, composed by both $Q$ and $\chi$. Different from ordinary composite Higgs models, the Higgs potential is not only from top and gauge loop corrections but also from the masses of preon $Q$. With this extra contribution, electroweak symmetry breaking can be realized differently and the correlation between the mass of Higgs and top partners is weakened.

To keep the Higgs potential from the top sector finite, we impose maximal symmetry in this model. Since the maximal symmetry can be realized in two different ways, we study its EWSB in two cases, the ordinary and minimal  maximal symmetric CHMs. In the first case, even the Higgs potential from the top and gauge sector is sensitive to composite resonance mass scales, the Higgs mass is only sensitive to the mass scale difference between composite vector mesons and top partners. So the composite partners of top and gauge bosons can be as heavy as possible, even around the cut-off $\sim 4\pi f$, for successful EWSB just at the cost of high fine tuning. While the top partner mass is around scale $f$ for light Higgs in ordinary CHMs without preon mass contributions no matter how to tune the parameters. However, since $\eta$ mass is related with the top partner mass if it is only from preon masses, positive $\eta$ mass square imposes the upper bounds on top partner mass, such as $M < 1.6$ TeV for $\xi=0.05$ and $m_\eta^2 >0$. But the mass of $\eta$ can also be generated from the hidden interactions which only break $\eta$ shift symmetry. If this contribution is dominant in $\eta$ mass, the correlation between the mass of top partners and $\eta$ is destructed and thus the composite top partners can be heavy arbitrarily (generally it should be smaller than cut-off scale) without any constraints. In the minimal maximal symmetric case, the Higgs  quartic from the top sector is suppressed at quartic order in top Yukawa coupling and is not sensitive to top partner scale $M_f$. However, if without preon mass contributions, the Higgs mass is always too light even $M_f$ is around cut-off scale, which always results in double tuning $\sim 95/\xi$. While the extra Higgs potential from preon mass can enhance the Higgs quartic so the Higgs mass can be heavy enough for several TeV top partners with $\xi=0.1$. Meanwhile, the tuning is significantly suppressed as low as minimal $\sim 1/\xi$ for $M>1.5$ TeV.
The $\eta$ mass is also insensitive to $M_f$ so it can be heavy enough to avoid the bounds $m_\eta > m_h/2$ and is almost fixed for fixed $\xi$ if its mass is only from preon mass sector (around $300$ GeV for $\xi =0.1$).

The partial compositeness predicts an extra $U(1)_\sigma$ pNGB $\sigma$ in this model. Since it contains both the freedoms of $Q$ and $\chi$, it can interact with SM EW and QCD gauge bosons through Wess-Zumino-Witten terms which is determined by UV completions. Especially its branch ratio into different gauge boson pairs can reveal the UV theory. This singlet can be resonance produced through gluon fusion and decay into gauge boson pairs. This can be the typical phenomenology of this kind of model at LHC and we derive the bounds of $\sigma$ mass for different $G_{HC}$ gauge group.
\section*{Acknowledgements}
 J.S. is supported by the National Natural Science Foundation of China
(NSFC) under Grants No.11947302, No.11690022, No.11851302, No.11675243,
and No.11761141011, and by the Strategic Priority Research Program of the
Chinese Academy of Sciences under Grants No.XDB21010200 and No.XDB23000000. T.M. is supported by the United States-Israel Binational Science Foundation~(BSF) (NSF-BSF program Grant No. 2018683) and the Azrieli Foundation.

\appendix

\section{Top partners and form factors in top effective Lagrangia}
\label{app:formula}
For the top partners $\Psi_{5,1}$ in the Lagrangian (\ref{eq:mix}), their explicit embedding in representation $\mathbf{6}$ of $SU(4)$ are
\begin{align}\label{eq:psi5}
\Psi_{5L} &= \left( \begin{array}{cc}
\frac{1}{2} T_5 i \sigma^2  & \frac{1}{\sqrt{2}} Q  \\
- \frac{1}{\sqrt{2}} Q^T  & \frac{1}{2} T_5 i \sigma^2
\end{array} \right) ,   \quad Q=\left( \begin{array}{cc}
T & X_{5/3}\\
B & X_{2/3}
\end{array} \right) \nonumber\\
\Psi_{1L}& = \left( \begin{array}{cc}
\frac{1}{2} T_1 i \sigma^2  &0 \\
0 & -\frac{1}{2} T_1  i \sigma^2
\end{array} \right)  \nonumber\\
\Psi_{5R}^c &= \left( \begin{array}{cc}
\frac{1}{2} T_5^c i \sigma^2  & \frac{1}{\sqrt{2}} Q^c  \\
- \frac{1}{\sqrt{2}} Q^{cT}  & \frac{1}{2} T_5^c i \sigma^2
\end{array} \right) ,   \: Q^c=\left( \begin{array}{cc}
-X^c_{2/3} & B^c \\
X^c_{5/3} & -T^c
\end{array} \right) \nonumber\\
\Psi_1^c& = \left( \begin{array}{cc}
\frac{1}{2} T_1^c i \sigma^2  &0 \\
0 & -\frac{1}{2} T_1^c  i \sigma^2
\end{array} \right)
\end{align}
The form factors in Eq.~(\ref{eq:effective_SU4}) and masses of top partners in Eq.~(\ref{eq:mtsim}) in an ordinary maximal symmetric model can be expressed as
\begin{align} \label{eq:form_factor}
\Pi_0 ^q  &=1-\frac{\lambda_L ^2 f^2}{p^2 - M^2}, \Pi_0 ^t  =1-\frac{\lambda^2 _R f^2}{p^2 - M^2},  M_1 ^t = \frac{\lambda_L \lambda_R f^2 M }{M^2 -p^2}\nonumber\\
M_{T_1}&=\sqrt{f^2\lambda_L^2+M^2},\quad M_{T_2}=\sqrt{f^2\lambda_R^2+M^2}.
\end{align}
The form factors in Eq.~(\ref{eq:effective_SU4_1}) and masses of top partners in Eq.~(\ref{eq:mMtopmass}) in the minimal maximal symmetric model can be expressed as
\begin{align}\label{eq:form_factor_mM}
&\Pi_0 ^q=1-\frac{\lambda_L ^2 f^2}{p^2 - M^2}, \Pi_0 ^t  =1-\frac{4\lambda^2 _R f^2}{p^2 - M^2}, M_1 ^t = \frac{\lambda_L \lambda_R f^2 M }{p^2 -M^2} \nonumber \\
&M_{T_1}=\sqrt{f^2\lambda_L^2+M^2},\quad M_{T_2}=\sqrt{4f^2\lambda_R^2+M^2}.
\end{align}

\section{Gauge sector}
\label{app:gauge}
According to the hidden local symmetry, the vector resonances $\rho_\mu$  transform non-linearly, while the axial resonances $a_\mu$ transform homogeneously, under a global $SU(4)$ transformation $ {\bf g}$,
\bea \label{eq:vector_resonaces}
\rho_\mu &=&  \rho^a _\mu T^a, \quad  \rho_\mu \to  {\bf h} \rho_\mu {\bf h} ^\dagger  + \frac{i}{g_\rho} {\bf h } \partial_\mu {\bf h}^\dagger \nonumber \\
a_\mu &=& a^{\hat{a}} _\mu T^{\hat{a}},  \quad   a_\mu \to  {\bf h} a_\mu {\bf  h }^\dagger,
\eea
where ${\bf h} ={\bf h}({\bf g},\pi^{\hat{a}})$ is the nonlinearly realised $Sp(4)$ element.
So at leading order in derivatives, the general Lagrangian allowed by Eq.(\ref{eq:vector_resonaces}) is
\begin{align}
\mathcal{L}_\rho &= -\frac{1}{2} \mbox{Tr}[\rho_{\mu \nu} \rho^{\mu \nu} ] +f_\rho ^2 \mbox{Tr}[(g_\rho \rho_\mu -E_\mu ^a T^a ) ^2]  \nonumber \\
\mathcal{L}_a &=  -\frac{1}{2} \mbox{Tr}[a_{\mu \nu} a^{\mu \nu} ] +\frac{f_a ^2}{\Delta^2} \mbox{Tr}[(g_a a_\mu -\Delta d_\mu ^{\hat{a}} T^{\hat{a}} )^2 ],
\end{align}
where $i U^\dagger D_\mu U =d_\mu ^{\hat{a}} T^{\hat{a}} +E_\mu ^a T^a$,  $\rho_{\mu \nu}  =\partial_\mu \rho_\nu -\partial_\nu \rho_\mu -i g_\rho [\rho_\mu, \rho_\nu]$, $a_{\mu \nu}=\bigtriangledown_\mu a_\nu -\bigtriangledown_\nu a_\mu$ and $\bigtriangledown_\mu =\partial_\mu -i E_\mu ^a T^a$.
After integrating out the heavy resonances at tree level, the $SU(4)$ invariant Lagrangian, at quadratic order in the gauge fields and in momentum space, is
\begin{align}
\mathcal{L} ^\text{eff} &= \frac{P_t^{\mu \nu}}{2}  \left( \Pi_0(p^2)  \mbox{Tr}[A_\mu A_{\nu}] - p^2( W^a _\mu W^a _\nu +B_\mu B_\nu)  \right. \nonumber \\
 &+ \frac{\Pi_1(p^2)}{4} \mbox{Tr}[ ( A_\mu \Sigma +\Sigma A_\mu^T   ) (A_\nu \Sigma +\Sigma A_\nu^T)^\dagger ]  \big),
\end{align}
where $A_\mu =g W_\mu ^a T^a_L +g^\prime B_\mu T_R ^3$, $P_t ^{\mu \nu} = g^{\mu \nu} - p^\mu p^\nu/p^2$ is the projector on transverse field configurations and $\Pi_{0,1}$ are form factors. From above Lagrangian, we get the most general effective Lagrangian for gauge bosons with explicit dependence on the Higgs field:
\begin{align*}
\mathcal{L}^\text{eff} &= \frac{P_t^{\mu \nu}}{2}\bigg( g^2 \Pi_0 ^W   W^a _\mu  W^a _\nu +g^{\prime 2} \Pi_0 ^B B_\mu B_\nu\\
&+ g^2\Pi_1 \frac{h^2}{h^2+\eta^2} \frac{s^2}{4} (W_\mu ^1 W_\nu^1 +W_\mu ^2 W_\nu ^2     )  \\
    &+\left.  \Pi_1 \frac{h^2}{h^2+\eta^2} \frac{s^2}{4} (g^\prime B_\mu - g W_\mu ^3 )(g^\prime B_\nu - g W_\nu ^3 )  \right),
\end{align*}
 where $s=\sin (\sqrt{h^2 +\eta^2} /f)$,
 \bea
  \Pi_0 ^W &=&  -\frac{p^2}{g^2} +  p^2 \frac{f_\rho ^2 }{p^2 -m_\rho ^2}, \;
 \Pi_0 ^B  = \Pi_0 ^W(g \to g^\prime),  \nonumber \\
  \Pi_1 &=& f^2 + 2 p^2\left( \frac{f_a ^2}{p^2 -m_a ^2 }   -\frac{f_\rho ^2}{p^2 - m_\rho ^2} \right).
 \eea
Here we define the mass parameters
 \bea
 m_\rho ^2 = f_\rho ^2 g_\rho ^2, \quad m_a ^2 = \frac{f_a ^2 g_a ^2 }{\Delta ^2}.
 \eea
So it is easy to get the Higgs potential at one-loop level by integrating out the gauge fields and going to Euclidean momenta space,
\begin{align} \label{eq:g_potential}
 &V_g(h) = \frac{3}{2} \int \frac{d^4 p_E}{(2\pi)^4} \left(2 \mbox{log}\big[ \Pi_0 ^W + \Pi_1 \frac{h^2}{h^2+\eta^2}  \frac{s ^2}{4} \big] \right.  \nonumber \\
   &+ \left. \mbox{log}\big[\Pi_0 ^B \Pi_0 ^W +\Pi_1 \frac{h^2}{h^2+\eta^2} \frac{s ^2}{4}(\Pi_0 ^B +\Pi_0 ^W ) \big]   \right ).
\end{align}

\section{A mechanism to producing heavy $\eta$}
\label{app:eta_mass}
In gauge and top sector, these interactions are $U(1)_\eta$ invariant so the SM fields loops do not contribute to $\eta$ potential. To produce a heavy $\eta$ while preserving Higgs and $\sigma$ shift symmetry, we introduce an electroweak singlet complex scalar $\phi$.  To break $\eta$ shift symmetry, we suppose it is embedded in ${\bf 6}$ representation of $SU(4)$ in the form
\bea
\Phi =\frac{\phi}{2} \left( \begin{array}{cc}
  i\sigma_2 &  {\bf 0} \\
{\bf 0} &  i\sigma_2  \\
\end{array}  \right).
\eea
So its general couplings to the pNGBs are given by
\bea
\mathcal{L}_\phi &= & \partial_\mu \phi^\dagger \partial^\mu \phi - m_\phi^2 \phi^\dagger \phi  -  y_\phi  f^2 \mbox{Tr}[\Phi \Sigma^\dagger] \mbox{Tr}[\Phi^\dagger \Sigma] \nonumber \\
& =&\partial_\mu \phi^\dagger \partial^\mu \phi -m_\phi^2 \phi^\dagger \phi -4 y_\phi  f^2  \frac{\eta^2 }{\pi_Q^2} \sin^2 \left( \frac{\pi_Q}{f} \right)   \phi^\dagger \phi.
\eea
The pNGB potential at one $\phi$ loop level is
\bea \label{eq:Veta}
V_\eta \simeq \frac{y_\phi f^2 C_\phi}{(4\pi)^2} \frac{\eta^2 }{\pi_Q^2} \sin^2 \left( \frac{\pi_Q}{f} \right) \Lambda ^2_Q,
\eea
where $\Lambda_Q \sim 4\pi f$ is the condense scale of preon $Q$ and $C_\phi$ is order one constant. Generally, $y_\phi C_\phi$ can be positive so $\eta$ becomes massive from the scalar loop. Its mass is naturally at $\mathcal{O}(f)$ so $\eta$ can be heavy enough to survive experimental bounds without any fine tuning.

\section{Mass of $\sigma$ from $\chi$ sector}
\label{app:sigma_mass}
The $\chi$ preon mass can explicitly break the shift symmetry of $\sigma$ and thus can contribute to $\sigma$ mass. The gauge invariant mass term of preon $\chi$ can be aligned with condensation $\Sigma_{\chi 0}$,
\bea
\mathcal{L}_{mass}^\chi=m_\chi  \chi^T _{l} \Sigma_{\chi0} ^{lm} \chi_{m} + h.c. \;,
\eea
where $m_\chi$ is the preon mass.
Notice that we can  formally keep the $SU(6)$ invariance by assigning the following  transformation rules to the mass matrix
\beq
 \Sigma_{\chi0}  \rightarrow g_\chi^* \Sigma_{\chi0} g_\chi^\dagger,
\eeq
where $ g_\chi \in SU(6)$.
Similarly, $\sigma$ potential from $\chi$ masses can be easily derived according to the global symmetry
\bea \label{eq:Vmchi}
V_m^\chi  &=& - C_\chi f_6 ^3 m_\chi \mbox{Tr }[\Sigma_{\chi 0 }.\Sigma_\chi ]+h.c. \nonumber \\
    & =& - 12C_\chi m_\chi f_6 ^3 \cos \left( \frac{2 \sigma \sin \phi}{\sqrt{6} f_6} \right),
\eea
where  $C_{\chi} \sim \langle \chi \chi \rangle /{(4\pi)^2 f_6^3}$ is the form factor related to strong dynamics~\cite{Galloway:2010bp}. We can also read the $\sigma$ mass from Eq.~(\ref{eq:Vmchi}),
\bea
\label{eq:sigmass}
m_{\sigma}^\chi
& = &  2 \sin \phi \sqrt{2 C_{\chi} m_\chi f_6 }  .
\eea


\end{document}